\documentclass{svjour3}
\usepackage[applemac]{inputenc}
\usepackage[a4paper, inner=3cm, outer=3cm, top=3cm, bottom=3cm]{geometry}
\usepackage{footmisc}

\usepackage{amscd,bbm}
\usepackage{natbib}
\usepackage{mathrsfs}
\usepackage{hyperref}
\usepackage[all]{hypcap}
\usepackage{braket}
\journalname{}
\begin{document}
\title{What is a physical theory? Philosophers do have an answer by distinguishing two forms of empirical adequacy}
\institute{Instituto de Investigaciones Filosóficas, Universidad Nacional Autónoma de México, Circuito, Mario de La Cueva s/n, Ciudad Universitaria, Coyoacán, 04510, Mexico City, Mexico. Email: jorge.manero@filosoficas.unam.mx. ORCID: 0000000345885132. Area of specialisation: Philosophy of science, metaphysics of science and philosophy of physics.}
\date{}
\author{Jorge Manero}
\maketitle
\begin{abstract} 
What is a physical theory? Although this intriguing question has been addressed from many different perspectives, some physicists and philosophers of physics have implicitly or explicitly embraced a philosophically-neutral definition of a  
physical theory, independently of the philosophical position endorsed with respect to it. Considering some objections against this view, I shall argue that the most appropriate definition of a physical theory already presupposes some commitments shared by a philosophical position associated with scientific realism. As we shall see, what physical theories and scientific realist positions have in common is the commitment of satisfying a non-factive notion of empirical adequacy, whilst a factive notion of empirical adequacy shall be solely associated with scientific realism. Based on this factive/non-factive distinction, we shall finally present a case study in physics in order to show that a well-known foundational problem associated with the concept of \emph{manifest time} can be dissolved once the non-factive notion of empirical adequacy (as opposed to factive one) is endorsed.  
\end{abstract}
\section{Introduction}
What is a physical theory? Although this intriguing question has been addressed from many different perspectives, one may classify the resulting answers into one of the following two camps: the philosophical approach associated with philosophy of science and the foundational approach associated with physics or philosophy of physics. This two-fold classification has focused its attention on how the question itself is interpreted, either as a \emph{normative} problem that attempts to make a rational reconstruction of physical theories according to the way they \emph{should be} ---as represented by philosophers through the use of their favorite meta-linguistic frameworks--- or as a \emph{descriptive} problem aimed at describing the way these theories really \emph{are} ---as presented by practicing scientists in the context of the actual practice of physics. 
\subsection{A two-fold classification of views towards theories}
As regards the philosophical approach, one may mention four contrasting, non-exhaustive philosophical positions that have something to say with respect to science in general and physical theories in particular: positivism, instrumentalism, antirealism and realism.\footnote{These philosophical positions may embrace more or less empiricist, pragmatist or rationalist commitments (or perhaps a mix of these epistemological temperaments).} The standard way each of these positions have answered the above question (or at least driven its discussion) has been amply focused on how should we generally \emph{represent} scientific theories as self-standing objects of study capable of doing fruitful and useful things, such as discovering, exploring, controlling, describing, predicting, explaining or simply reasoning about the phenomena. For example, many scholars of more or less positivist temperament have tried to represent scientific theories through the so-called `syntactic’ or `received’ view, whilst there are more instrumentalists, antirealists and realists who prefer to represent them through the so-called `semantic’ view or even both, provided a dual picture of scientific theories is endorsed.\footnote{As is well-known by philosophers of science (disregarding, of course, many subtle variations among each view), the mere task of the syntactic view, on the one hand, is to represent the features and components of a given theory in terms of a more or less rigid, formal and linguistic mold (i.e., a set of logico-linguistic propositions closed under logical deduction) capable of discerning its constitutive ingredients and giving them meaning only via their correspondence with the empirical world ––in a similar way as we do with words when reading a text paragraph. On the other hand, the task of the semantic view is to represent this theory through a more flexible, non-fixed, extra-linguistic representational scheme ––normally associated with a family of interacting models–– considering the complex, structured, and dynamical dimension of scientific practice. As noticed recently, this distinction is neither a necessary divide nor a sharp dichotomy as there are alternative views, such as the \emph{dual approach} advocated by \citep{frigg2022}, the main purpose of which is to interpret theories as having common features of both  syntactic and semantic views, provided some of their problems can be addressed by integrating them into a consistent whole.}

As regards the foundational approach, people of more `tough-minded’ temperament have been interested in supplying an approximate definition of a physical theory based on its purely physical aspects and not on its philosophical interpretation. Contrary to the philosophical approach, the central point of this camp is to investigate the anatomy of all successful physical theories by identifying the basic conceptual categories used by physicists throughout the history of physics (e.g., the concepts of matter and motion, the spacetime structure, the process of mathematisation, etc.), some of which might not be applicable to other scientific domains. Although this way of defining a physical theory has been implicitly or intuitionally embraced by many physicists, the best-known example of expressing such implicit or intuitive ideas in terms of a canonical definition can be found in \citep{maudlin2019}. In this contribution, Tim Maudlin tries to come up with a supposedly ``neutral” (although it is taken as \emph{cum grano salis}) canonical definition of a physical theory based on a sharp distinction between his ``philosophically disinterested” way of defining a physical theory and the philosophical positions someone could endorse with respect to such a theory and others.  
\subsection{Objections}
Notwithstanding the influential consequences of both approaches, some objections may be raised against them. 

Firstly, there fairly reasons to think that the above distinction drawn by the foundational camp cannot be sustained as supposedly ``neutral” canonical definitions are incapable of transcending the philosophical standpoint in which they are based. As corroborated by the `Weltanschauungen Analysis' of scientific theories in general---amply discussed in \citep[119-221]{suppe1977}---, any human construction, such as a physical theory, cannot be defined in a \emph{brute}-canonical way as it depends on an immanent \emph{conceptual framework} from which we are looking the world, ourselves and our own material and conceptual inventions. As a result, the way we conceive of a physical theory is highly dependent on the state of knowledge it is embedded, together with the aims, interests and commitments one would have at that moment with respect science in general and physics in particular. These aims, interests and commitments may range over normative claims associated with value-laden assumptions of epistemic or non-epistemic kind ––underlying the complex axiological structure of physical theories–– to descriptive claims associated with methodological, epistemological and metaphysical presuppositions ––shedding light on how these theories are discovered, designed, implemented and justified.  Under these circumstances, the most appropriate way to define a physical theory seems to stand within our favorite conceptual framework, provided we are transparent and clear with respect to it in the sense of making explicit the aims, interests and commitments endorsed by such a framework. 

Secondly, this task of transparency does not only compromise us with a philosophical analysis of a physical theory from the standpoint of a clear conceptual framework, but also with a metaphilosophical analysis of the philosophical position in terms of which the alleged physical theory is defined. That is, we should be explicit with respect to the aims, interests and commitments endorsed by the `second-order’ conceptual framework from which we are trying to define the respective philosophical position. As a consequence, the philosophical approach is not immune to the objection raised against the foundational camp. As it is the case for the latter, the former also needs to recognise that any philosophical position, such as positivism, instrumentalism, antirealism or realism, is a human construction analysed from the perspective of an immanent conceptual framework. This means that the way we conceive of a philosophical position is highly dependent on the state of knowledge, together with the aims, interests and commitments one would have at that moment with respect to the task of practicing philosophy. 

In addition to the last objection, one might follow the lines of \citep{frenchcosta,laudan2007} and argue that the philosophical approach is conflating \emph{our best reconstruction of theories}, according to our favorite representational framework at the meta-level, with the reification of this reconstruction: \emph{the way theories really are}. For example, one might think that the best way of characterising scientific theories (at the meta-level of the philosophy of science) is in terms of the semantic model-theoretic view, but this is not to say that theories are such models (at the level of science). We should not demand that scientists themselves adopt these kinds of formal, representational frameworks at the level of their practice. Actually, a similar objection came from \citep{quine1969} who not only recognised the above distinction but also argued against rational reconstructions and in favor of analysing how scientific theories are in fact developed as they appear in actual practice. By doing this, he was in fact laying the basis of the foundational approach towards theories. 
\subsection{The new philosophical approach}
Considering Quine's critique, but without collapsing to the foundational approach, some advocates of a new school arising from the philosophical camp (hereinafter the \emph{new philosophical approach}) have tried to answer the main question of this contribution (generalised to any scientific theory) by surpassing the normative task of choosing their favorite meta-level framework. On the contrary, they have engaged in two different (but related) descriptive tasks:\footnote{The distinction between functional and ontological characterisation of theories is taken from \citep[Ch. 14]{frigg2022} by extrapolating (without loss of generality) his analysis on models to scientific theories.} the \emph{functional} characterisation of scientific theories based on their constitutive roles played in science, on the one hand, and their \emph{ontological} characterisation based on what kinds of objects they are, on the other. 

As regards the ontological task, they have came to interpret scientific theories as \emph{being} material objects \citep[Ch. 2]{weisberg}; abstract artifacts with multiple ontological instantiations (discoveries) \citep{thomasson2020}; or even fictions without having any ontological role to play \citep[Ch. 4 y 6]{Barberousse2009,contessa2010,frigg2022,levy2019}. However, apart from the obvious underdetermination of these ontological characterisations by the theory, when it comes to interpret scientific theories in the particular context of physics, these characterisations seem to be highly dependent on the whole complex set of practices associated with this scientific domain, such as the physical concepts used, the mathematical language applied (in terms of which theories are presented), the experimental tools employed, etc. As a result, if the most complete and reasonable answer to our main concern is to be sought in the relevant practices directly, any ontological characterisation of scientific theories seem to be superfluous and dispensable. In other words, there is no need to increase our ontological baggage in order to identify and individuate a theory if it only gets its meaning through the associated practices. Along this way of thinking,\citep{french2020} has came to interpret scientific theories as being nothing (ontologically speaking), concluding that \emph{there are no such things as theories}, but merely scientific practices.

As regards the functional task, the most famous candidate among functional characterisations of theories is the special role played by them concerning the complex process of \emph{scientific representation}. The advantage of characterising a scientific theory in terms of this functional character is that there is an immense amount of literature that has been written addressing this issue. The problem, however, is that there is in fact no way of capturing the variety of scientific theories we encounter in scientific practice under one type of scientific representation \citep{frigg2022}. Indeed, as far as the representational roles of theories are concerned, the way they are defined is highly dependent on commitments associated with one or more philosophical positions. For example, when theories are characterised as representing some of their (unobservable) targets in ways that are inaccurate, this characterisation seems to be compatible with instrumentalism and some form of antirealism, whilst it is incompatible with scientific realism. Thus, considering that certain philosophical positions play a constitutive role in relation to the way theories are functionally interpreted, they should be unveiled and stated explicitly before any functional characterisation is articulated ---as corroborated by our previous observation that some transparency among scientific and philosophical theories is required at the philosophical and metaphilosophical level. This shall be the general task to be developed in this contribution. 
\subsection{Methodological assumptions and limitations}
Considering the relativised nature of the above functional characterisation within the new philosophical approach, the only way to define a scientific theory based on a particular and transparent notion of scientific representation is to constrain the scope of this contribution to a particular level of analysis and to a fixed domain of scientific and philosophical knowledge.

As regards the level of analysis, I shall locate the `first-order’ conceptual framework from which a scientific theory is investigated at the philosophical level, whilst I shall locate the `second-order’ conceptual framework from which a philosophical position is investigated at the level of metaphilosophy. As regards the domain, I shall restrict the respective philosophical analysis within the limits of \emph{physical science}, whilst the metaphilosophical analysis shall be confined within the limits of \emph{scientific realism} (to be canonically defined in this work on the basis of standard literature). 

As a result, we shall try to establish the most appropriate definition of a physical theory in relation to scientific realism, provided an appropriate characterisation of the latter is developed. Although some ontological characterisations of physical theories, such as the artifactual or the fictional might share commitments associated with scientific realism, we shall concentrate on the functional roles played in such theories and on their relation with this philosophical thesis.

Another important constraint to be considered (for sake of space) is that both physical theories and scientific realist positions shall be interpreted from the perspective of a conceptual framework solely associated with constitutive commitments (e.g., the satisfaction of a well-defined notion of empirical adequacy within the current state of knowledge, as we shall see), as opposed to regulative aims or interests of achieving something along the process of theorising (e.g., generality, unity, permanence, etc.). In other words, instead of asking what physical theories and scientific realist positions look for, we shall focus on the following two research questions:  

\begin{enumerate}
\item[(1)] What is a physical theory? \\
\item[(2)] What is scientific realism?
\end{enumerate}

As a result, we shall only be interested on the constitutive commitments (as opposed to the regulative aims and interests) endorsed by the conceptual framework from which physical theories and scientific realist positions are analysed. 
\subsection{The main objective}
Given the above two questions and the respective constraints, the \emph{purpose of this contribution} is to defend a generalised  version of Maudlin’s canonical definition of a physical theory from the perspective of the new philosophical approach ---as opposed to his philosophically-disinterested foundational approach. As required by a functional characterisation of physical theories, I shall restrict the metaphilosophical analysis to a transparent and clear philosophical position associated with scientific realism. Consequently, I shall argue that Maudlin's definition already presupposes some commitments shared with this philosophical thesis, the identification of which shall be achieved by distinguishing two different notions of empirical adequacy: 

\begin{itemize}
\item[(I)] A factive notion of empirical adequacy (hereinafter, FEA).\\
\item[(II)] A non-factive notion of empirical adequacy (hereinafter, NFEA). 
\end{itemize}

Based on a more general definition identified with NFEA (slightly different from Maudlin's), I shall conclude that what physical theories and scientific realist positions have in common is the commitment of satisfying NFEA, whilst the FEA shall be solely associated with scientific realism. This means that both (1) and (2) questions cannot be disassociated as the answer to the former (i.e., NFEA) includes the answer to latter (i.e., FEA) as a necessary (although non-sufficient) condition. 

For illustrative purposes, we shall present a case study in physics in order to show that a well-known foundational problem associated with the concept of \emph{manifest time} can be dissolved once NFEA (as opposed to FEA) is endorsed. According to our main argument, this means that this problem can be dissolved once we obtain a clear idea of what a physical theory is on the basis of NFEA and certain commitments associated with (but not reduced to) scientific realism.  

The methodological procedure shall be as follows: In section \ref{sec2}, I shall present the standard view of scientific realism followed by an elaborated, alternative definition of this philosophical position based on the distinction between FEA and NFEA. In section \ref{sec3}, I shall present Maudlin’s definition of a physical theory followed by its association with NFEA (resulting in a comparative analysis with our alternative definition of scientific realism). Finally, in section \ref{sec4} I shall present the above case study followed by some concluding remarks in section \ref{sec5}. 
\section{What is scientific realism?}\label{sec2}
\subsection{The standard view of scientific realism}
Let us first express the standard definition of scientific realism. According to the literature in the field \citep{van,psillos,real}, this philosophical thesis may be generally defined in terms of the conjunction of three necessary components:\footnote{These authors define scientific realism in this way, regardless of whether or not they are advocates of this view.} 

In addition to the well-known \emph{metaphysical component} of scientific realism ---associated with the existence and the mind-independent status of the external world---, the \emph{semantic component} is, put simply, the condition establishing how scientific theories are to be interpreted and what they are supposed to be saying with respect to the world without assuming that they are correct, whilst the \emph{epistemic component} states, in addition to this semantic condition, that what these theories say about the world is approximately correct. Let us say something more about each of these components. 

Leaving aside the metaphysical component (to be taken for granted as a necessary condition), the semantic component was initially introduced by \citep{feigl,horwich,van} and later defined by \citep[p. 10]{psillos} as an interpretative commitment to read scientific theories literally. Considering \citep{van}'s reading of a literal interpretation, such a semantic commitment is associated with the idea that theoretical assertions are capable of being expressed in a form that they could have (at least potentially) truth values and that certain conditions to obtain those values should be specified according to an appropriate regimentation of the theory. Although it is important to stress that without the semantic component there cannot be an epistemic component, a literal interpretation should not be confused with the stronger condition (associated with the epistemic component) of determining the truth values of these assertions (i.e., to assert whether or not the alleged truth conditions obtain) \citep[p. 10]{psillos}. An illustrative example, well-known in the philosophical literature, is the metaphysical claim that ``God exists''. As expected, atheists and Christian devotees disagree on the true value assigned to this claim, whilst they agree on the literal interpretation of this claim, meaning that when they utter that claim they are assigning the same extension to it (and, therefore, the same concepts to the terms ``God'' and ``Existence''). 

Furthermore, the semantic component stipulates that theoretical terms, such as ``electron waves'', have the sole function of referring to putative entities of a possible world in which the theory is the case. This means that this component should inform us what the world would be like from the immanent standpoint of this theory as if everything said by it were supposedly correct ---as if electrons waves were existing entities behaving in the way the theory says, although it might not be the actual case. In contrast, the epistemic component states, in addition to this mere semantic correspondence, that theoretical terms approximately refer to the real entities that constitute the actual world. 

Note that the semantic component is compatible with anti-realism, namely, the view that denies one (or all) of the three components of scientific realism, as defined below. This is because this component is a necessary but non-sufficient condition for scientific realism. For example, van Fraassen's antirealist position known as constructive empiricism assumes the semantic component but denies the epistemic one. Indeed, he insists that ``not every philosophical position concerning science which insists on a literal construal of the language of science is a realist position \citep[p. 11]{van}''. To arrive at this conclusion, he divides the anti-realists into two sorts, ``The first sort holds that science is or aims to be true, properly (but not literally) construed. The second holds that the language of science should be literally construed, but its theories need not be true to be good.'' Based on this distinction, he makes clear his position by saying ``The anti-realism I shall advocate belongs to the second sort \citep[p. 9-10]{van}”. As a result, the semantic component of scientific realism can be endorsed irrespective of whether we are realists or antirealists.

In the following subsections, we shall expand the standard definition of scientific realism by identifying additional criteria against which our best scientific theories are to be analysed.  
\subsection{Scrutinising the semantic component: Non-factive empirical adequacy}
According to our own reading of the above definition, we propose to identify the semantic component of scientific realism with a criterion I shall name as \emph{non-factive empirical adequacy}. This criterion is defined as: 
\begin{quote}
The capability of our best scientific theories of \emph{explaining} by means of a \emph{clear ontology} the \emph{phenomena} associated with the possible world (not necessarily the actual one) where such theories are the case. 
\end{quote}
Let us try to clarify the notions of a `clear ontology', `explanation’ and `phenomena’ involved in this definition. 
\subsubsection{Ontological clarity}
In brief, by ontological clarity I mean the semantic virtue of a theory of being clear with respect to what it says about the possible world in which it is the case. This means that, according to this condition, the referents of theoretical terms (together with predicates' extensions) must be assigned by the theory through a well-defined ontology that has no considerable traces of semantic ambiguity. In this way, ontological clarity is a possessed virtue of some scientific theories to be unveiled by a philosophical analysis, in the sense that there are theories that satisfy (or do not satisfy) such a virtue independently from the philosopher's will.  Let us make three important remarks concerning this virtue.

Firstly, the definition of a theoretical virtue, such as ontological clarity, is not immune to further concerns associated with its normative and regulative nature: How much clarity one needs? How ontological clarity is interpreted among different theories? As is the case for any kind of theoretical virtue, ontological clarity is immanent and contextual in the sense that it is highly dependent on the kind of ontology posited by a given scientific theory, apart from the fact that there is no unique way to specify in any particular situation the precise conditions for this virtue to be satisfied. For example, some people committed with other theoretical virtues, such as simplicity, expect that some theory satisfies ontological clarity if the associated ontology is only constituted by a minimum set of theoretical posits indispensable to explain and predict the relevant phenomena, whilst others think that this theory should go deeper and characterize the metaphysical natures of such posits at the price of positing a more robust and a less simple ontology. This normative character of ontological clarity is a reflection of the fact that the way scientific realism is defined depends on a `second-order' conceptual framework where regulative and normative claims are involved, such as simplicity or (metaphysical) strength. Although these claims form part of an important element to be considered, it is important to reiterate that in this contribution our focus relies on the constitutive dimension of scientific theories. 

Secondly, considering the above `constitutive' definition of ontological clarity, let us stress the association of this criterion with the semantic component of scientific realism. As corroborated by a proper reading of \citep{van}, ontological clarity is, in part, a necessary condition for the semantic component because it is a condition that makes literal interpretations possible. In van Fraassen words: “a literal construal of a theory can elaborate by identifying what theoretical terms designate”, provided that “on a literal construal, the apparent statements of science really are statements, capable of being true or false \citep[p. 10]{van}.” Furthermore, ontological clarity is a virtue that makes the semantic component different from the epistemic component because a literal construal is not related to “our epistemic attitudes towards theories, nor to the aim we pursue in constructing theories, but only to the correct understanding of what a theory says” \citep[p. 11]{van}.  

Finally, one might claim that the ontology of a scientific theory involved in the definition of ontological clarity not only includes the referents of theoretical terms (e.g., electrons, quarks, genes, and so on), but also certain one-place or many-place predicates satisfied by the extensions of those terms, such as their properties (e.g., charge, quark color, locus, and so on) and, specially, the corresponding laws of the theory. Although endorsed by many (but not all) scholars, the above claim presupposes the metaphysical assumption that the laws of nature are existing entities pertaining to the scientific world depicted by the theory in question. As such, they can be interpreted in the primitive, `theological' or fundamental sense that they govern, determine or even produce the events of the world, enforcing the behavior of the referents of theoretical terms by means of necessitarian relations or primitive universals \citep{dretske1977,armstrong1983,tooley,maudlin}; but also they can be interpreted in the dispositional sense in which they prescribe the disposition of the alleged referents to behave in certain way (conceived of as governing powers) \citep{shoemaker,mumford,bird}. However, we should take this claim with caution as it depends on a non-exclusive ontological view with respect to laws of nature. In contrast to this view, we can interpret laws in an eliminativist way as having no reason to believe they exist \citep{cartwright1983,van1989,giere1999}; alternatively, we can interpret them in a descriptive `humean' sense such that they are true generalizations that appropriately summarise the patterns displayed by a mosaic of events or matter of fact \citep{lewis,loewer,callender,bhogal}. From this point of view, the laws of nature are non-fundamental, constructive elements arising from the fundamental ontology of the theory and, according to a particular humean account, from certain normative and regulative standards, such as simplicity and informative strength, associated with the way the theory develops.   
\subsubsection{The nomological notion of the phenomena}
According to NFEA, phenomena are not brute empirical facts of the actual world waiting to be explained by a conscious human being capable of accessing that part of the world through her best scientific theory. On the contrary, they are putative, theoretical explanations and predictions confined within a given scientific theory incapable of transcending the scope of the possible world in which such a theory is the case (where such a possible world is not necessarily the actual one). Thus, when talking about phenomena in the context of Newtonian gravitational theory we are not referring to regularities in the sky observed through the naked eye; instead, we have in mind something like immensely large particle aggregates in elliptical motion explained and predicted by Newton's universal gravitational law, the worldview of which is no more that absolute space, external forces, point-wise particles without any external human mind capable of confirming the theory with an ingenious telescope. As such, phenomena are disassociated from the epistemological context of justification as they constrain the scope of the `empirical compartment' of a theory (to be clarified below), irrespective of whether or not such phenomena are confirmed by available empirical facts in the actual world. More specifically, they serve as a reservoir of potential theoretical explanations and predictions to be actually confirmed in case there were someone capable of perceiving the actual world in some way or another. This means that in context of NFEA the phenomenological domain depends on what the theory allows to be in principle explained and predicted, independent of how we humans or any other agent assess that theory on the basis of their perceptive capabilities. Let us analyse this in more detail. 

Many philosophers of empiricist spirit think of the phenomena of a scientific theory as constituted by empirical facts. Although these facts are normally interpreted as brute, self-standing elements that ultimately arise from a mind-independent world ---as far as the metaphysical component of scientific realism is assumed---, they appear to us as concrete, available sense data, the structure of which is the result of a worldly-human process mediated by our perception system ---sometimes called \emph{unaided sense perception}. For example, the primary set of empirical facts that scientists usually read from their complicated measuring devices before any kind of systematisation process takes place is ultimately displayed in the form of some sensitive effects occurring in the manifested world, such as a ‘pointer’ heading in a certain direction. Thus, when the phenomena are interpreted as empirical facts, it seems that these phenomena are displayed in a form accessible through the human senses.

Moreover, in order to say something about empirical facts, they should be represented in some way or another. The way these facts are represented depends, of course, on an immanent representational framework capable of organising and conceptualising the available sense data. The nature of this representational framework depends, in turn, on the kind of knowledge that arises from such facts.

When ordinary people represent empirical facts ---provided ordinary people are human agents with a well-functioning unaided sense perception---, the respective framework is associated with what is called \emph{perceptual knowledge}.\footnote{For a recent comprehensive description of perceptual knowledge and its relation to scientific knowledge see \citep{teller2021}.} This framework organizes the available sense data in terms of certain useful and stock-in-trade concepts confined within the scope of everyday language. Although these concepts are associated with an imperfect, idealised framework incapable of capturing the nature of unaided sense perception in its full complexity, they are successfully used in everyday thinking and ordinary affairs of practical kind. For example, the motion of a $1m^{3}$-volume, grey, iron sphere in free fall is perceived by the eye as a temporal succession of extended positions possessed by a rigid object in physical space, and this sense perception is approximately represented by familiar concepts, such as ``rigid object of spherical shape'', ``volume'', ``grey color'', ``iron'' and ``temporal succession''. Although these concepts do not represent the perceptual situation in exact terms (as any real object of this kind is not perfectly spherical, grey and iron-made), they suffice to indicate us, for all practical purposes, that we are in the face of an acknowledgeable empirical fact.

However, the way scientific theories represent empirical facts depends on a completely different representational framework to that of everyday language. Coming back to the previous example, the iron sphere in free fall is represented by Newtonian theory (considering the appropriate idealisations and approximations) as a large configuration of point-wise particles (forming a composed spherical object of about $1m^{3}$ size), describing multiple integral curves of a vector field in a mathematical Euclidean space. As learnt from Susan Stebbing's objection against Eddington's famous two-tables tale, everyday language and scientific language should not be mixed; on the contrary, the framework associated with a scientific theory involves a conceptualisation of empirical facts that cannot be corrupted by the concepts used in everyday language. 

Thus, once a given empirical fact is represented by a scientific theory independently of everyday language, some set of `empirical data' is obtained. This resulting representation, constrained within the epistemic limits of the theory, are theoretical elements located in some region of its conceptual space. Where exactly they are located depends on the way we represent, at the metaphilosophical level, the complex structure of scientific theories. From the point of view of the orthodox version of the syntactic framework, empirical data are represented as synthetic propositions disassociated from the formal structure of the theory (i.e., analytical propositions), whist from the point of view of the (also orthodox) semantic approach, they form a family of data models (i.e., substructures contained in and related to other structures of the theory) that make `observable' propositions of the theory true. Irrespective of the choice made among these alternative frameworks, they share the premise that within the limits of a certain theory there is a set of empirical data, distinguished from the rest of its elements, having the potential quality of representing the realm of empirical facts, in case such a theory were assessed through sense detection or experimentation. 

But if we transcend the context of justification beyond the mere task of designing experiments to confirm the available explanations and predictions, and scientific theories are interpreted, in turn, as self-standing, conceptual constructions that take on a life of their own ---through, for example, working hypotheses about a possible world whose true values are assumed to be irrelevant---, the question is whether or not some set of empirical data can still be distinguished from the rest of the elements contained in the theory. The answer, according to NFEA, is that there is in fact a distinguished set, which we shall call \emph{nomological phenomena} (from here to the end of this section, let us just call it `phenomena'). 

For example, Newtonian mechanics can be clearly formulated as a theory of interacting point-wise particles and external forces exerted upon them. But if these tiny, idealised particles interact with each together by creating complex mereological configurations sising at the scale of about $10^{-1}m$, and these configurations interact with other configurations of the same size scale, one may obtain phenomena having the potential of being qualitatively identical (at least approximately) to macroscopic objects, such as tables and chairs, whose macroscopic properties can be observed through the human senses. Note, however, that this mapping identity between a set of microscopic particles and macroscopic tables and chairs, directly observable by the naked eye, is an anthropocentric measure or reference external to the theory that is put by hand in order to confirm (though the mediation of our visual perceptual system) Newton's theory predictions. Thus, it seems that if we are not interested in confirming Newton's theory, the phenomena associated with this theory should be merely interpreted as theoretical elements, such as configurations of particles, whose aggregate properties obey certain effective principles or laws associated with rigid bodies (i.e., Euler's laws of motion) different from (although compatible with) those that govern the individual particles. 

The problem with NFEA's answer is what makes the aforementioned distinction possible in case there are no available empirical facts arising from the actual world to be represented by the corresponding phenomena. Going back to the previous example, the problem is to identify without anything external to Newton's theory the distinguishing qualities associated with the phenomena (i.e., configurations of particles and their aggregate properties governed by Euler's laws of motion) that are not shared by the rest of the elements contained in the theory (e.g., individual particles, forces, Newton's laws, etc.). 

One well-known answer associated with constructivist empiricism is to the bite the bullet and postulate by fiat a fundamental epistemological distinction between observable and unobservable elements of the theory. The idea is to say that there are two kinds of claims and terms posited by scientific theories: observable claims together with observable terms (e.g., tables and chairs), on the one hand, and unobservable claims together with unobservable terms (e.g., electrons and quarks), on the other. The problem with this response is that \citep{van} is implicitly distinguishing observables from unobservables by means of an anthropocentric criterion that gives meaning to the observability of certain phenomena. In fact, according to him, what is observable is what is in principle observable by us (throughout  our past, present and future), in the sense that observables claims and terms are empirical facts that appear to us as concrete, available sense data. In this way, \citep{van} ends up embracing something external to the theory (i.e., the way humans perceive the world), a situation which we have denied from the start in the context of NFEA. 

Another answer is to deny that there is a fundamental epistemological distinction between observable and unobservable elements of the theory; in contrast, there is a distinction but of modal nature (similar to \citep{chackra}' distinction between detection and auxiliary properties), in the sense that observable claims/terms (as opposed to unobservables) are theoretical elements with at least two potentially-based propensities that can be realised depending on certain conditions imposed by the actual world (and, therefore, external to the theory): either being qualitatively identical to some empirical facts ---in case observable claims of the theory are true and observable terms are referring terms--- or being qualitatively different to them ---in case those claims are false and observable terms do not refer. Although the alleged propensities serve as distinguishing qualities by means of which one clearly differentiates observable from unobservable claims/terms, as long as we stay within the context of NFEA (where something external to the theory is neglected), these propensities shall remain \emph{in potentia} without being actually realised in the way just described. 

There is, however, one objection that can be raised against this response. The idea is that nothing within the theory allows us to introduce the alleged metaphysical propensities associated with the way observables claims/terms are distinguished and characterised. Since those propensities are introduced by contrasting the body of theoretical elements within the theory with the available empirical facts in the actual world, we are introducing fixed metaphysical commitments (and increasing the metaphysical baggage of the theory, in turn) based on empirical data that could change as our experimental and detection capabilities are improved. In other words, what distinguishes the observable claims/terms of the theory from the rest of its elements would be fixed, and this means that the theory would be incapable of bringing some unobservable theoretical elements to the space of observables, contrary to the evidence we have from the history of science.  

Considering the above critical comments, we can suggest a better response if we focus our attention to the actual process through which scientific theories are discovered. This response is based on the observation that any theoretical construction, such as a scientific theory, ultimately originates from sense data, even if it is later considered as an immanent body of working hypotheses and empirical data incapable of transcending the scope of the possible world in which such a theoretical construction is the case. As nicely claimed by Whitehead, ``The true method of discovery is like the flight of an aeroplane. It starts from the ground of particular observation; it makes a flight in the thin air of imaginative generalisation; and it again lands for renewed observation rendered acute by rational interpretation'' \citep[p. 5]{white1}. In this way, at a given early stage during the development of scientific theories, there are some empirical facts available from which certain working hypotheses arise. For example, by a process of abductive reasoning one may be able to perceive certain evidence and try to explain that evidence in terms of some possible conjecture or hypothesis. Once the resulting hypotheses are settled, one may dispense from the empirical facts that served to produce them and only retain the theoretical counterparts of such facts, which we have called the phenomena (i.e., theoretical representations of worldly empirical facts). However, the way the theory has being constructed, according to this reasoning, involves a complex inner structure (i.e., either a hierarchical or model-based structure depending on the metaphilosophical framework endorsed) where the phenomena of the theory are distinguished from the corresponding hypotheses and the rest of theoretical elements. The available empirical facts in the actual world is like the plastic mold used to produce some shapes when cooking some cookies: once their shapes are formed, one can dispense from the mold and the cookies retain their form. In this case, the preserved form is precisely the structure that codifies the distinction between the phenomena and the rest of the theory's elements. Thus, it seems that the phenomena can be identified and distinguished from other elements of scientific theories based on the way these theories are discovered. 

Let us come back to the previous Newtonian example to explain the proposed response. Newton's theory was discovered, in part, through regular observations of a multiplicity of empirical facts, such as Galileo's famous experiments or the precise experimental data obtained by Tycho Brahe. Once the laws of Newton were established, together with other theoretical elements (e.g., absolute space, Newtonian forces, Newtonian particles, etc.), the theory was completed as we approximately know it today. More specifically, it turned out to be a selfstanding body of working hypotheses that could be disassociated from the above empirical facts and generate a multiplicity of nomological possibilities independent of these facts, such as idealised friction-free models of inertial motion. Note that from the immanent standpoint of the completed theory, the phenomena that initially represented the available empirical facts as observed through our senses, such as the real motion of a 10-cm-diameter iron ball moving down an inclined wood plane, were (strictly speaking) no more than configurations of particles whose aggregate properties obey Euler's laws of motion. However, the `mold' left by the Newtonian explanation of the actual iron ball's motion in terms of the ontology of the theory (and, therefore, the distinction between microscopic particles and macroscopic rigid bodies) did not evaporate; rather, it was preserved to the extent that the physics of rigid bodies emerged as an effective theory of large configurations of particles, each of which obeys Newton's laws. 

This alternative response opens the possibility of having a distinction which is not epistemologically fundamental, as van Fraassen claims, but does not have to increase the metaphysical baggage of the theory at the price of drawing a once-and-for-all distinction, as the modal response seems to suggest. Rather, it is a relativised distinction, defined with respect to a certain stage of knowledge, that can change as scientific theories develop. From our point of view, what ultimately distinguishes the phenomena from the rest of the elements of a scientific theory are the multiple nomological possibilities generated by a family of working hypotheses compatible with some set of empirical facts available at a certain stage of knowledge. As a result, we endorse a nomological conception of the phenomena, the central role of which is to focus on the multiple possibilities of a theory to generate novel predictions and explanations independent of the epistemological task of confirming such predictions and explanations though the mediation of our sense perceptions. 

This view is reinforced by the observation that scientific theories are more complex structures than sense-based machineries whose sole function is to represent empirical facts, as perceived by the human senses, in terms of a theoretical calculus. On the contrary, scientific theories frequently posit entities that generate and predict novel phenomena, some of which transcend the human-based domain of sense data. Thus, the phenomena may comprise not only empirical data that have been experimentally detected, but also scientifically possible data that cannot be (or have not been) displayed in a form accessible through the human senses. Illustrative examples are idealised models, such as  friction-free inertial motion of point-wise particles (already mentioned above), and thought experiments seen throughout the history of science: without inquiring into the extensive literature on this issue, some of these experiments are phenomena that have not been experimentally detected, but are in principle detectable by  any possible perceptual system capable to be described as any other constitutive element of a theory (whose properties supervene on some properties of such a theory). It follows from this last example that in the context of a given theory nomological phenomena are in principle detectable by any agent (human, non-human, or whatever entity or property) living in a possible world where such a theory is the case. 
\subsubsection{Purported explanation}
In his famous first chapter of \emph{Process and Reality}, Alfred N. Whitehead defines `speculative philosophy' as 
\begin{quote}
``[...] the endeavor to frame a coherent, logical, necessary system of general ideas in terms of which every element of our experience can be interpreted.'' \citep[p. 3]{white1}
\end{quote}
provided this system is applicable and adequate with respect to its interpretation, in the sense that ``some items of experience are thus interpretable'' and that ``there are no items of experience incapable of such interpretation.'' \citep[p. 3]{white1}. 

Inspired by pragmatist ideas of his time, Whitehead acknowledges that such an endeavor is never final, but it is an adventure in ``the clarification of thought [...] in which even partial success has importance'' \citep[p. 9]{white1}. As a result, he is in some way committing himself to a fallible (but still objective) stance towards any system of philosophical ideas, where some pragmatic criteria associated with this system (i.e., coherence, logical consistency, generality, necessity, applicability and adequacy) should be satisfied, irrespective of whether or not it leads to objective true. According to this stance, these criteria are pragmatic as they are associated with immanent, contextual and normative standards defined with respect to a given stage of knowledge and confined within the limits of the empirical facts that are epistemically accessible at that stage (what he understands as the ``system's sphere of validity''). This is clear when Whitehead says that the alleged system ``[...] is a matrix from which true propositions applicable to particular circumstances can be derived.'' \citep[p. 8]{white1}. Contrary to objective truth, the aim of any philosophical theory is, according to him, the mere process of understanding, which is 
\begin{quote}
``[...] never a completed static state of mind. It always bears the character of a process of penetration, incomplete and partial. ...Of course in a sense, there is a completion. But it is a completion presupposing relation to some given undefined environment, imposing a perspective and awaiting exploration.'' \citep[p. 43]{white3}. 
\end{quote}
Furthermore, he insists that we should refrain from committing ourselves to what he calls the `fallacy of misplaced concreteness', which happens when one mistakes an abstract and working hypothesis about the way things are for concrete reality. 

Considering these well-known observations, we would like to focus on Whitehead's pragmatic notion of `experience' (as part of his definition of a speculative scheme), as it retains the spirit of what we define as `phenomena' in our definition of NFEA. Once this correspondence is established, we shall see that his notion of `interpretation' is directly associated with our notion of `explanation' (to be defined here in more precise terms), except from the fact that in the latter case the associated scheme is contextualised to scientific theories ---as if it were the worldview depicted by a scientific theory. As a consequence of this association, we shall be able to motivate our definition of NFEA in terms of a reconceptualisation of Whitehead's speculative scheme in the context of scientific theories. 

As already said above, Whitehead sustains that the root of any system of philosophical ideas arises from a restricted set of available empirical facts. However, once this system is settled and takes on a life of its own, its functional role merely consists in the ``clarification of thought'' by constructing a theoretical edifice of working hypotheses and empirical data governed by the rules of logic and other pragmatic criteria. Analogous to the already established distinction between phenomena and other theoretical elements of a theory, Whitehead argues that this theoretical edifice has its rational side and its empirical side. The rational side is expressed by the regulative criteria of coherence and logical consistency, whilst the empirical side is expressed by those of applicability and adequacy. Furthermore, the relativised nature of such a distinction is also illustrated by Whitehead's insistence that the rational and empirical sides are bound together by virtue of the fact that any coherent, logical and necessary system of general ideas located in the rational side should be applicable and adequate with respect to the empirical side, provided ``the texture of observed experience, as illustrating the philosophic scheme, is such that all related experience must exhibit the same texture.'' \citep[p. 4]{white1}. This quotation presupposes an interpretative-laden notion of experience that points towards the impossibility of perceiving brute empirical facts disassociated from the theoretical framework used to represent them; instead, all ``elements of experience'' conceptualize and represent the available empirical facts, provided they are immanently associated with a theory in an way that cannot be abstracted from it. In his own words:
\begin{quote}
 ``[...] there are no brute, self-contained matters of fact, capable of being understood apart from interpretation as an element in a system. [...] Thus the understanding of the immediate brute fact requires its metaphysical interpretation as an item in a world with some systematic relation to it. When thought comes upon the scene, it finds the interpretations as matters of practice. Philosophy does not initiate interpretations. Its search for a rationalistic scheme is the search for more adequate criticism, and for more adequate justification, of the interpretations which we perforce employ. Our habitual experience is a complex of failure and success in the enterprise of interpretation. If we desire a record of uninterpreted experience, we must ask a stone to record its autobiography.'' \citep[p. 14-15]{white1}
\end{quote}
Considering this observation in the context of scientific theories, it follows that Whitehead's notion of experience can be nicely associated with the notion of nomological phenomena defined above. Along Whitehead's lines, we are interpreting phenomena as distinguished elements of a theory capable of representing ---in case such a theory is confirmed--- some confined set of empirical facts in a way that illustrates and exhibits a possible world in which such a theory is the case. As he put it, ``Every scientific memoir in its record of the `facts' is shot through and through with interpretation.'' \citep[p. 15]{white1}. 

Granted this association, let us finally demonstrate how Whitehead's notion of `interpretation' directly motivates the notion of `explanation' involved in the definition of NFEA. In so doing, let us remind the reader that the way scientific theories say something about the available empirical facts is by representing them in terms of some set of phenomena immanently incorporated into its theoretical edifice. However, when the worldview depicted by a scientific theory is regarded as an instantiation of a system of philosophical ideas concerning the world ---provided some sort of naturalistic criterion towards philosophy is endorsed---, it is not hard to see that this notion of representation can be directly linked to Whitehead's notion of interpretation, and since the latter can be conceived as explanatory in the ontological sense, it is therefore linked to a notion of metaphysical explanation. Let us inquire into some details of this claim.

By `interpretation' Whitehead means ``that everything of which we are conscious, as enjoyed, perceived, willed, or thought, shall have the character of a particular instance of the general scheme'' \citep[p. 3]{white1}. Considering that ``everything of which we are conscious, as enjoyed, perceived, willed, or thought'' is associated with theoretical or perceptual representations of available empirical facts (and not with facts themselves), this quote seems to suggest that the task of interpreting these representations (which we have called `the phenomena') can be associated with an ontological, albeit putative, notion of explanation (hereinafter, \emph{purported explanation}), the role of which is to account for these phenomena, as explanandum, in terms of more fundamental principles or hypotheses, as explananda, immanently confined within the theoretical edifice of the theory in question. In more precise terms, this notion of purported explanation has to do with the metaphysical problem of accounting for how the the phenomena within a scientific theory are recovered from the fundamental ontology of that theory. This means that with the help of certain theoretical resources ---and perhaps an ingenious metaphysical toolbox (e.g., supervenience, grounding and/or metaphysical dependence)--- the empirical counterpart of the theory must be cashed out in terms of its underlying fundamental ontology, whose entities and qualities pertain to the possible world in which such a theory is the case. Let us finally make three important remarks concerning this purported notion of explanation. 

Firstly, purported explanations involve a non-factive, pragmatic notion of explanation associated with the semantic component of scientific realism. To see why, let us remind the reader that this component requires that scientific theories should tell a literal story about the possible world in which these theories are the case. It follows from this requirement that this story is purportedly explanatory in the sense that it explains the phenomena in a way that some theoretical virtues are exhibited (e.g., ontological clarity, coherence, logical consistency, generality, necessity, applicability and adequacy) although it could be wrong. In other words, purported (as opposed to factive) explanations of the phenomena are putative propositions capable of ensuring predictive success on account of the ontology of the theory, without being compromised with their truth values. 

Secondly, purported explanations are interpretative-dependent. This is clear when one associates this notion of explanation with the metaphysical problem of accounting for the phenomena in terms of the ontology of the theory. Associated in this way, the ontological clarity criterion is a necessary condition for having purported explanations as the ontology of the theory must be clearly specified before indicating the way this ontology recovers the observed phenomena. If a precise ontology is not specified, there is no way to tell a literal story of how the scientific world is according to this theory. 

Finally, our third remark has to do with the universality ascribed to this purported notion of explanation. As argued by Whitehead, ``[...] the philosophic scheme should be `necessary', in the sense of bearing in itself its own warrant of universality throughout all experience'' \citep[p. 4]{white1}. And based on this claim, he later concludes that `[...] The metaphysical first principles [associated with this scheme] can never fail of exemplification'' \citep[p. 4]{white1}. In this way, the notion of purported explanation involves a criterion of universality according to which the nomological phenomena explained by a theory is like a hologram, where each point of it exemplifies the whole theory and its fundamental laws and principles. For example, when planetary motions are interpreted by Newton's worldview, they are conceived as phenomena governed by the same laws and principles to that of an idealised ball in friction-free inertial motion. In this way, we say that the idealised ball's behavior exemplifies the underlying laws and principles of the Newtonian universe. 
\subsection{Scrutinising the epistemic component: Factive empirical adequacy}
According to the standard definition of scientific realism expressed above, we propose to identify its epistemic component with another criterion known as \emph{factive empirical adequacy}. This criterion is defined as: 
\begin{quote}
The capability of our best scientific theories of \emph{explaining} by means of a clear ontology the available \emph{empirical facts} associated with the actual world (considering the respective idealisations and approximations). 
\end{quote}
As was the case for the non-factive notion of empirical adequacy, the latter definition is also prone to further concerns. However, considering that the same notion of `ontological clarity' is a necessary condition for both the semantic and epistemic components of scientific realism, we only need to clarify the notions of `explanation’ and `empirical facts’ involved in this definition. 
\subsubsection{Empirical facts}
As opposed to nomological phenomena, empirical facts are constitutive elements of the actual world displayed in a form accessible through the human senses. Considering that a comprehensive definition of `empirical facts' has been already established (as it was required to arrive at the notion of nomological phenomena), we only wish to mention the most important elements associated with it. Indeed, by empirical facts we mean those parts of the actual world which are epistemically accessible by some living agent capable of perceiving that world in some way or another. As such, empirical facts are necessarily associated with the epistemological context of justification as they form the epistemic basis of our scientific theories and rational beliefs. This means that, contrary to NFEA, the phenomenological domain does not depend on what a theory allows to be in principle explained and predicted, but on the way the actual world is and how we humans or any other agent assess that theory on the basis of their perceptive and detection capabilities.
\subsubsection{Factive explanation}
As opposed to purported explanations defined above, FEA involves a factive notion of explanation the role of which to account for the phenomena by means of true propositions ensuring predictive success on account of what the actual world is really like. Defined in this way, factive explanations are associated with the epistemic component of scientific realism (as opposed to purported explanations, which are associated with the semantic one). Indeed, once we assume that the literal story told by scientific theories is approximately true, the notion of explanation changes from being pragmatic to being factual in the sense that this story involves true propositions, all of which are (part of) the reason why some empirical facts occur. 

Furthermore, contrasted to purported explanations, this factive notion of explanation is neither interpretative-dependent nor universal. One the one hand, it is not dependent on the interpretation of the theory because it accounts for the available empirical facts in terms of true propositions concerning what the actual world is really like. As such, the ontology of the theory in terms of which empirical phenomena are explained is constituted by entities and qualities pertaining to the actual world, provided there is only one true and literal interpretation. On the other hand, it is not universal in the strict sense as it is not necessary with respect to all the phenomena associated with the theory in question; on the contrary, factive explanations contingently depend on what empirical facts are capable of being detected through unaided sense perception. For example, predictions obtained by the theory that cannot be detected by sense perception, such as friction-free inertial motion or certain thought experiments, are excluded from what we define as factive explanations. This means that factive explanations are confined to that which communicates with knowable, immediate matter of fact and excludes, therefore, scientifically possible data that cannot be (or have not been) displayed in a form accessible through the human senses. It follows that in the context of a given theory factive explanations account for empirical facts that are detectable by any agent with unaided sense perception living in the actual world. 
\section{What is a physical theory?}\label{sec3}
\subsection{The foundational view}
We should come back to the introduction and ask again the following question: What is a physical theory? In one of his recent contributions, \citep{maudlin2018} argues that one cannot truly understand a physical theory if one is not clear with respect to what kind of physical world it presents beyond the mere instrumental task of generating predictions. In so doing, he proposes an approximate canonical definition of a physical theory where ``the precise ontological commitments of a theory are made manifest and unambiguous in a disciplined way'' \citep[2]{maudlin2018}. 

As already said in the introduction, this way of defining a physical theory has its roots on (although it is not necessarily linked to) a foundational approach, the central point of which is to investigate the anatomy and the conceptual categories of all successful physical theories that have been constructed throughout the history of physics. Indeed, after a brief revision of the history of the concept of ``matter in motion'', Maudlin concludes that physics can be canonically defined as the most general account of what there is and what it does, at the fundamental level of physical reality. In his more recent book, this definition is expressed as follows: 
\begin{quote}
``A physical theory should clearly and forthrightly address two fundamental questions: what there is, and what it does. The answer to the first question is provided by the ontology of the theory, and the answer to the second by its dynamics. The ontology should have a sharp mathematical description, and the dynamics should be implemented by precise equations describing how the ontology will, or might, evolve.'' \citep[11]{maudlin2019}
\end{quote}
In certain physical contexts (those related to ``matter in motion''), this definition involves ``the description of local beables [i.e., entities with well-defined location in space-time] and whatever else is required to account for their trajectories through space-time'' \citep[3]{maudlin2018}. Although this description does not exhaust all physical possibilities (as there can be space-time theories without trajectories or theories that are not defined in space-time), Maudlin argues that there are, at least, four basic postulated categories of what counts as constitutive of a physical theory: local beables, non-local beables, space-time structure and dynamical equations (principles and laws).\footnote{NB: there are other elements discussed by him that enter into the definition of a physical theory, such as the role of mathematisation and derivative ontology. These elements shall be discussed above when addressing the relation between Maudlin's foundational view and our definition of NFEA.} Note, however, that the above categories are established with respect to the current stage of knowledge, and since they arise from a historical revision concerning the notion of ``matter in motion'' that have shaped the anatomy of all ever-proposed successful physical theories, Maudlin's canonical definition is to be regarded as tentative, waiting to be modified or expanded as physics develops. 

Of course, all the categories mentioned above can be scrutinised in a more robust way by characterising their metaphysical natures. In this spirit, one might attempt to draw metaphysical inferences from scientific theories by inquiring into the nature of space-time, the identity and metaphysical priority of local and non-local beables, the fundamental status of the laws of Nature and many other metaphysical debates prevailing in the philosophical literature. However, considering that a necessary condition for drawing these metaphysical inferences from theories is to be sufficiently clear about the ontology posited by them, the point of Maudlin's canonical definition is to strike the right balance between a `metaphysically shallow' and an `anti-instrumentalist' attitude towards physics and identify the common and necessary conceptual categories from which all these inferences follow.\footnote{In a recent workshop, Maudlin took `anti-instrumentalism' as the position that faithfully captures his definition of a physical theory.} This presupposes a distinction between the semantic task of achieving certain level of transparency with respect to the ontological commitments of a theory and the epistemological task of drawing certain metaphysical conclusions from these commitments concerning what exactly exists in this world.  

Having decided to address the semantic (as opposed to the epistemological) task, \citep{maudlin2019} goes deeper in his conceptual analysis and scrutinizes this semantic-epistemic distinction by differentiating between his ``philosophically disinterested” way of defining a physical theory and the philosophical positions someone could endorse with respect to such a theory and others. This is clear when he accuses the systematic misuse of philosophical terms, such as `realism' or `antirealism', having a precise meaning in the context of philosophy of science, but which are completely unfamiliar to most physicists. According to him, ``physical theories are neither realist nor antirealist. That is, as we used to say, a category mistake. It is a person’s attitude toward a physical theory that is either realist or antirealist'' \citep[12]{maudlin2019}. Based on this differentiation, he argues that any question concerning the realism-antirrealism debate ``is a question addressed by epistemology and confirmation theory''. After giving some examples from contemporary literature, he finally concludes that this kind of philosophical questions are not (and should not be) addressed by someone doing foundations of physics. In his own words: ``Whether some person’s attitude toward the theory is one of scientific realism or not is neither here nor there. If I had my druthers, ``realist'' and ``anti-realist'' would be banned from these foundational discussions'' \citep[13]{maudlin2019}. 
 \subsection{Our view of physical theories and scientific realism}
In the following lines, we shall argue that, contrary to Maudlin's foundational view, his canonical definition of a physical theory already presupposes some philosophical commitments shared by scientific realism. In fact, opposed to the above sharp distinction between the task of doing philosophy and the task of doing foundations, we shall see that there are in fact philosophical questions addressed by the foundational camp as well as there are foundational issues addressed by the philosophical camp. 

Let us remind the reader that two different notions of empirical adequacy can be distinguished: A factive notion of empirical adequacy (FEA) and a non-factive notion of empirical adequacy (NFEA). Considering this distinction, we shall see that both physical theories and scientific realist positions share the commitment of satisfying NFEA, whilst scientific realism only satisfies FEA. This shall imply that the canonical definition of a physical theory involves certain commitments included in the definition of scientific realism, provided these commitments are to be regarded as necessary (although non-sufficient) conditions for scientific realism. 
\subsubsection{What is really scientific realism?}
As argued in the last section, the semantic and epistemic components of scientific realism are associated with certain criteria involved in the definition of NFEA and FEA, respectively. For sake of precision, however, let us again explain this association.

On the one hand, the semantic component is constitutively identified with NFEA for the following reasons. According to this component, scientific theories should be interpreted literally, provided they are clear with respect to what they say about the possible world in which they are the case (even though such a possible world might differ from the actual one). 
Defined as such, this semantic commitment can be directly linked to certain capability of our best scientific theories of purportedly explaining by means of a clear ontology the nomological phenomena associated with a possible world in which they are the case. The important point to be emphasised from this link is that, apart from the well-established notion of ontological clarity, the notions of explanation and phenomena involved in this definition of NFEA are interpreted in a particular way, leading to their compatibility with the semantic component of scientific realism. Indeed, we are talking here of nomological phenomena and purported explanations, both of which are defined as non-factive, pragmatic notions that are immanently associated with the theory in question, independently of its epistemic justification. As regards purported explanations, this means that they are not compromised with the question of truth but with satisfying certain pragmatic and regulative criteria, such as ontological clarity, coherence, logical consistency, generality, necessity, applicability and adequacy.  

On the other hand, the epistemic component is constitutively identified with FEA for other reasons. According to this component, scientific theories should not only be interpreted literally but also factually, in the sense that their posited theoretical terms refer to existing entities of the actual world, provided what these theories say about this world is approximately true. Defined as such, this epistemic commitment can be directly linked to certain capability of our best scientific theories of factually explaining by means of a clear ontology the available empirical facts associated with the actual world. As above, the important point to be emphasised from this link is that, apart from the well-established notion of ontological clarity, the notions of explanation and phenomena involved in this definition of FEA are interpreted in a particular way, leading to their compatibility with the epistemic component of scientific realism. Indeed, we are talking here of empirical facts (as opposed to phenomena) and factive (as opposed to purported) explanations. In contrast to NFEA, both notions are associated with the context of epistemic justification and, therefore, they are compromised with the question of truth. In particular, empirical facts are regarded as brute, contingent facts of the actual world epistemically accessible by means of unaided sense perception, whilst factive explanations are true propositions accounting for the available empirical facts. 

Let us note, however, that the commitment to the epistemic component of scientific realism involves the necessary commitment to the semantic one. Indeed, although literal interpretations are not the same as true interpretations, it is necessary to have a literal, clear and empirically adequate interpretation of a theory just to specify its true values. Thus, the semantic component is a necessary (but non-sufficient) condition for the epistemic component. Furthermore, since fully-fledged scientific realism necessarily assumes the metaphysical, semantic and epistemic components, it follows from the above observation that endorsing the epistemic component is the same as endorsing fully-fledged scientific realism. In the context of the previous discussion, this means that if the epistemic component is identified with FEA, it turns out that fully-fledged scientific realism can also be identified with FEA. 

Considering that the semantic component of scientific realism is constitutively identified with NFEA and fully-fledged scientific realism with FEA, let us see how these identifications serve us to locate Maudlin's definition of a physical theory in the realism-antirealism context. As we shall see, this philosophical contextualisation opposes Maudlin's foundational view; instead, it endorses an alternative view in which the task of doing foundations and the task of doing philosophy are intertwined.
\subsubsection{The philosophical dimension of physical theories} 
It is clear from the above discussion that Maudlin's canonical definition of a physical theory already presupposes some philosophical commitments shared by the semantic component of scientific realism. As we shall see, these commitments are precisely those involved in the definition of NFEA.  

Let us recall that, according to Maudlin, physics can be canonically defined as the most general account of what there is and what it does, at the fundamental level of physical reality. This definition is, of course, related to the semantic component of scientific realism as it embraces a commitment to ontological clarity and other criteria associated with a literal interpretation that takes scientific theories at face-value. In more precise terms, behind Mauldin's definition there is a commitment to clearly posit entities (i.e., local and non-local beables) that behave according to some preestablished rules (i.e., probabilistic or deterministic physical principles and laws), the physical form of which are trajectories traveling through space-time. Granted this observation, let us see how Maudlin's definition is associated with our definition of NFEA.

First of all, the criterion of ontological clarity is clearly endorsed by Maudlin's definition as it embraces the requirement of any physical theory to posit a clear and well-defined ontology in terms of which the referents of theoretical terms (together with predicates' extensions) are assigned. One might argue that Maudlin is assuming in his definition not only the requirement of being clear with respect to the ontology of the theory, but also the task of settling its dynamics through the postulation of a set of fundamental laws in terms of which the ontology evolves. If this were the case, one might argue that there is in fact some difference between his definition and NFEA. However, although not explicitly expressed in \citep{maudlin2018,maudlin2019}, we have already mentioned above (when characterising the notion of ontological clarity in the realist context) that Maudlin's ontological view towards natural laws involves the assumption that the ontology of the theory not only comprises the referents of theoretical terms but also certain many-place predicates satisfied by the extensions of those terms, such as its fundamental laws. In this way, the fundamental principles and laws can be said to be part of the ontology of the theory, rendering the association between his definition and NFEA well-suited (at least, as far as the ontological clarity criterion is concerned).

Furthermore, Maudlin expands on the definition of ontological clarity in the context of physics by establishing the condition that any physical theory should provide a sharp mathematical description of the posited ontology, its dynamics and the corresponding space-time structure in which it is located. In his own words: ``In presenting a clear and precisely articulated physical theory, then, one ought to specify the fundamental physical ontology and how the fundamental physical ontology is to be represented mathematically.'' \citep[7]{maudlin2018}. This presupposes, of course, a (frequently ignored) distinction between the alleged physical entities and their mathematical representation. Granted this distinction, the way physical theories provide this mathematical description depends on what Maudlin calls `a commentary', which merely consists in relating by means of appropriate approximations and idealisations the mathematical representation to the represented physical entities. These approximations and idealisations are, in his own terminology, mathematical fictions employed for certain practical purposes associated with the alleged representation, such as calculating predictions. Considering that there is no canonical way for establishing such a relation, this commentary should make clear whether or not the mathematical objects representing  the physical system in question denote something in the physical world, and if they do, what kind of entities they are supposed to be denoting. In his own words, the alleged commentary clarifies which mathematical degrees of freedom correspond to the represented physical system and which rather are `gauge', in the sense that have no physical counterpart in the degrees of freedom of the represented system (i.e., they form surplus mathematical structure). 

Secondly, the phenomena associated with any physical theory is, according to Maudlin's definition, physical events occurring in space-time. Indeed, the phenomena in this case are not associated with empirical facts perceived through unaided sense perception but with space-time trajectories predicted by the physical theory. Of course, a single space-time trajectory is not sufficient to obtain the phenomena associated with any physical theory; rather, there must be a considerable number of space-time trajectories all of which compose a rigid space-time object of the order of about $10^{-1}$ m (in the spatial dimension), provided this object obeys certain effective principles and laws that equally apply in the classical and macroscopic domain of unaided sense perception. Of course, the way each these effective principles and laws arise from the dynamics of the multiple trajectories shall depend on the theory in question as it has to explain how they are recovered from actual laws and principles obeyed by the posited ontology. This is the requirement presupposed by a purported notion of explanation, as we shall see next. 

Thirdly, the notion of explanation presupposed by Maudlin's definition is associated with the capability of physical theories of purportedly explaining by means of a clear fundamental ontology the phenomena associated with complex compounds of space-time trajectories. These phenomena are, in his own terminology, derivative ontology, which mainly comprise physically-possible entities composed of more fundamental entities. As such, phenomena are physically real entities, according to the possible world in which the theory is the case, but they are not fundamental as they are definable in terms of the fundamental ontology and its dynamics. Furthermore, the effective principles and laws that apply to the phenomena are \emph{sui generis} in the sense that their predictive effectiveness are explained by means of fundamental laws. After all, fundamental laws do not influence the phenomena as such but the fundamental entities in terms of which these phenomena are composed. In more precise terms, there can be no fundamental physical laws that refer to composed entities whose qualities are identical to classical and macroscopic objects accessible through unaided sense perception; on the contrary, fundamental laws refer to trajectories in space-time, whilst effective laws refer to complex compounds of these trajectories. 

Let us finally propose a general definition of a physical theory based on this philosophical contextualisation of Maudlin's work but which broadens its scope in a way that can be applied to any theory of this kind.
\subsubsection{What is really a physical theory?} 
Considering the above link between Maudlin's definition of a physical theory and our definition of NFEA, let us note that  this link is not symmetric. Although Maudlin's definition implies NFEA, the converse is not true. This is because his definition of a physical theory does not exhaust all possibilities associated with all ever-known physical theories. As already said above, local beables are, according to him, associated with a concept of `matter in motion' whose main feature is to be located in regions of space-time in the form of well-defined trajectories. However, this is not entirely true for posited entities in general as there can be space-time entities without having trajectories (e.g., physical fields, distributions of mass, and so on), something which he surely acknowledges, or entities that are not defined in space-time (e.g., events and causal relations in causal set theory). 

Under these circumstances, instead of endorsing Maudlin's definition of a physical theory, we prefer to endorse the one associated with NFEA. The advantage of endorsing the latter is that it is a more general definition that embraces not only the above theoretical possibilities but that can be applied to any past, present and future physical theory capable of positing a clear and well-defined ontology in terms of which the phenomena associated with that theory are explained. Although the notions of ontological clarity, purported explanation and nomological phenomena involved in NFEA are dependent on the stage of knowledge associated with the theory, the definition of NFEA itself does not depend on such epistemic contingencies. On the contrary, it is an objective definition that applies to any physical theory that have been (and will be) developed, irrespective of any historical revision we could have with respect to already-established physical theories. 
\section{Case study}\label{sec4}
It is an undeniable fact that we humans conceive of time as that concept which embraces many features, such as the idea that it flows; that events come into being in a succession of nows; that these nows are special to our life as we believe that they define our immediate (but objetive) present standpoint with respect to the world; and that the past, already known and settled, is fundamentally different from the unknown future of possibilities. Known as \emph{manifest time} in the literature, this concept is not a brute empirical fact that we directly experience through unaided sense perception, but it is an useful and stock-in-trade representation of many complex aspects of the world that cannot scape from the veil of human prejudice as well as it is confined within the scope of ordinary life. In other words, it is ``a regimented common sense picture of the world [...] We arrive at manifest time no differently than we do many of our other models of the world, through a mix of innate endowment, experience, and cognitive inference'' \citep[2]{callender}. 

Physical science, on the other hand, has developed a completely different picture of time to that of its manifest counterpart. Known as \emph{physical time}, it is a concept whose definition depends, of course, on the particular physical theory in terms of which it is framed. For example, Newtonian physics interprets time as a privileged foliation in Euclidean spacetime inducing simultaneity surfaces of absolute space, whilst relativistic physics interprets time as a (minus-sign) coordinate of a Minkowski space-time. However, when it comes to the essential features of time shared by all ever-known physical theories, there seems to be a problematic mismatch between these and everyday temporal features. This mismatch is not surprising as we normally see throughout the history of physics that new scientific discoveries radically change the worldview associated with our regimented common sense representations of the world. Ultimately, any quick analysis into the nature of time among our most successful physical theories suggests us that manifest time does not seem to have a counterpart in physics. The problem of reconciling these two pictures of time is what philosophers of physics call \emph{the problem of time}. 

Defined in this way, one might think that this problem is merely an interesting puzzle to be solved by the mind of an ingenuous thinker without having any relevant philosophical consequences. However, we shall not interpret such a problem in this way; on the contrary, we shall focus on how and why scientific realism (or the notion of FEA, as defined above) poses a strong philosophical motivation to address it. As we shall see, this problem only makes sense if we endorse the notion of FEM, whilst it is dissolved if we only endorse NFEA. Considering that any physical theory has been defined according to NFEA, we shall conclude there is no problem of time to be solved by any physical theory. 

If our aim is to epistemically access the actual world through the lens of our best physical theories, it follows that any regimented common sense picture of the world, such as manifest time, is either wrong or non-fundamental. That is, from the standpoint of a realist attitude towards physical theories, manifest time can be `radically' interpreted as a false representation of something real whose nature can only be captured in terms of the fundamental notion of physical time; but also it can be `moderately' interpreted as an approximately true representation whose meaning can be reconceptualised in terms of the purely notion of physical time ---in a way that manifest time arises from a more fundamental representation of the actual world informed by our best physical theories. Considering both interpretations, we cannot scape from the unavoidable fate associated with the problem of time: a solution to it is required if we think our best physical theories truly represent the actual world. Indeed, we need to understand why we are wrong concerning our manifest conception of time or how this conception arises from physical time, provided certain fundamental features associated with the latter are unnoticed by creatures like us. 

Without loss of generality, let us focus on the `moderate' way to interpret the notion of manifest time and think of this notion as a true representation reconceptualised in terms of physical time. Granted this, we can see from the above observation that, from the scientific realist standpoint, the mere formulation of the problem of time, and therefore, the problem of accounting of manifest time in terms of physical time is grounded on the assumption that the latter is more fundamental than the former and that any physical theory should have the capability of explaining by means of a clear fundamental ontology the available empirical facts arising from that part of the actual world associated with the objective notion of time. Since the way these time-related facts can only be known by us through the mediation of our perceptual and representational capabilities, the problem of time can ultimately be expressed as the capability of a physical theory of explaining by means of a clear ontology the notion of manifest time. This means that the problem of time only makes sense within the epistemic context of justification as any explanation and prediction obtained by the fundamental ontology of the theory ---and by the notion of physical time, in particular--- can only be confirmed through the available time-related empirical facts associated with the actual world, provided these facts are immediately known to us by means of a sense-based representation associated with manifest time. In other words, the formulation of this problem depends on what the actual world is really like, apart from the possibility of epistemically accessing that world through the lens of our best physical theories.

Coming back to the main discussion of this contribution, I hope it is clear that the problem of time, as formulated above, is directly associated with the notion of FEA. That is, such a problem only makes sense if we assume the epistemic premise that any physical theory should have the capability of explaining by means of a clear ontology the available empirical facts associated with the actual world. However, note that we have defined a physical theory as a theoretical construction that should at least obey NFEA and this means that there is no need to satisfy FEA. Therefore, any physical theory does not need to solve the problem of time as this problem only makes sense within the context of FEA. In other words, if we only endorse NFEA, which is the definition of a physical theory as stated above, the problem of time is dissolved. Let us explain this in more detail.

As already argued, the distinction between NFEA and FEA is grounded on a subtle distinction between the possible world immanently associated with (true or false) physical theories and the actual world capable to be epistemically accessed through the lens of approximately true theories. In the context of the notion of time, this subtle distinction means that one thing is to construct a possible world in which a putative notion of physical time is the case and another thing is to represent the actual world in which a correct notion of physical time should be justified. The former (as opposed to the latter) case does not need to tell a story of how and why manifest time, which is a sense-based representation of that part of the actual world associated with objective time, arises from physical time. Since manifest time is highly dependent on the way the actual world is, apart from the way humans perceive and represent that world, it is independent from NFEA. According to our definition of a physical theory, one only needs to explain by means of a clear ontology (in terms of which physical time is framed) the phenomena associated with the possible world in which such a theory is the case. From this point of view, the time-related phenomena are not associated with manifest time but with theoretical inferences arising from the notion of physical time. As a consequence, there is in principle no problem of time to be solved by any physical theory. 
\section{Concluding remarks}\label{sec5}
In defining a physical theory there are at least two different approaches that have been analysed in this contribution: the standard and new philosophical camps associated with philosophy of science ---which depend on a philosophical attitude towards physical theories--- and the foundational camp associated with physics or philosophy of physics ---which embraces a philosophically-neutral attitude towards them. Although the latter camp has been useful to bring about a canonical definition of a physical theory, I have argued that the most appropriate definition of a physical theory, framed in terms of a functional (as opposed to ontological) characterisation, already presupposes some commitments shared by a particular philosophical position: scientific realism. In more precise terms, we have argued that physical theories and scientific realist positions have certain philosophical commitments in common associated with a non-factive notion of empirical adequacy as opposed to a factive notion of empirical adequacy solely associated with scientific realism. This argument was reinforced by an illustrative case study that appeals to the problem of time. As a consequence, we conclude that a physical theory is functionally defined in terms of the non-factive notion of empirical adequacy.
\section*{\emph{Funding}} I confirm that this work was supported by CONAHCYT under the scholarship ``Estancias posdoctorales por México 2022(1)'' [Grant number: 442599]. This funding source has no involvement in study design; in the collection, analysis and interpretation of data; in the writing of the report; and in the decision to submit the article for publication.

\end{document}